\documentclass[10pt,longbibliography,aps,prl,twocolumn,floats]{revtex4-2}
\usepackage{graphicx}
\usepackage{xcolor}
\usepackage{amsmath,bm}
\usepackage{amssymb}
\usepackage[colorlinks]{hyperref}

\DeclareMathOperator{\diag}{diag}

\begin{document}

\title{Exceptionally deficient topological square-root insulators}

\author{Subhajyoti Bid}
\affiliation{Department of Physics, Lancaster University, Lancaster, LA1 4YB, United Kingdom}
\author{Henning Schomerus}
\affiliation{Department of Physics, Lancaster University, Lancaster, LA1 4YB, United Kingdom}
\date{\today}
\begin{abstract}
One of the most surprising features of effectively non-Hermitian physical systems is their potential to exhibit a striking nonlinear response and fragility to small perturbations. This feature arises from spectral singularities known as exceptional points, whose realization in the spectrum typically requires fine-tuning of parameters. The design of such systems receives significant impetus from the recent conception of \emph{exceptional deficiency}, in which the entire energy spectrum is composed of exceptional points.
Here, we present a concrete and transparent mechanism that enforces exceptional deficiency through lattice sum rules in non-Hermitian topological square-root insulators. We identify the resulting dynamical signatures in 
static broadband amplification and non-Abelian adiabatic state amplification, differentiate between bulk and boundary effects, and
outline routes to implementation in 
physical platforms.

\end{abstract}

\maketitle

Effectively non-Hermitian physical systems can exhibit striking responses to small perturbations, a behavior rooted in spectral singularities known as exceptional points (EPs) \cite{kato,Dembowski2001,Heiss_2004,Berry2004,Heiss_2012,Miri2019}. At an EP, the eigenstates associated with the degenerate eigenvalue become identical, in contrast to the situation in Hermitian systems where the eigenstates form a basis. 
This eigenstate coalescence leads to 
dramatically altered static and dynamical responses to external  perturbations \cite{Trefethen99,jan2} and driving \cite{Schomerus2020,Hashemi2022} down to the quantum limit  \cite{Yoo:2011,takata2021observing,Simonson2022}, which can be exploited, for instance, for sensing  \cite{jansensor,Chen2017,HHW17,XLK19,Lai2019,Wiersig2020,KCE22}, lasing \cite{Peng2016,MZS16}, amplification \cite{Zhang2019,ZOE20}, and 
mode conversion \cite{XMJ16,DMB16}.
Realizing EPs typically requires fine-tuning, making their robust engineering a central challenge. The number of parameters required to obtain a single EP in the spectrum can be lowered by symmetries, such as parity-time symmetry in gain-loss balanced structures \cite{El-Ganainy2018}. 
The realization of higher--order EPs, in which the number of coalescing states is larger than two, has benefitted from the introduction of non-reciprocal and unidirectional coupling schemes,  which can be realized in topolectric \cite{helbig2020generalized,Zou2021}, mechanic \cite{brandenbourger2019non,ghatak2020observation,Zhang2021,veenstra2024non}, and quantum-optical \cite{liang2022dynamic} lattice systems as well as resonator-waveguide arrangements \cite{KGK23,SZM22}. 
Despite these advances, conventional EP implementations remain limited in scope: their functionality is typically restricted to narrow spectral regions near isolated EPs,  and their realization often hinges on delicate parameter tuning or symmetry constraints that fix the position of the EP in the spectrum. This poses a fundamental limitation for broadband applications or robust device design.
In light of this, the recently introduced concept of \textit{exceptional deficiency}  \cite{Zli2025}---in which every energy level of a system is an exceptional point---marks a significant conceptual breakthrough. It opens the door to broadband EP-enhanced phenomena and offers a new route toward practical non-Hermitian functionalities. 
The original proposal, implemented in active mechanical lattices,
achieves this through a tailored coupling between two subsystems with matching spectra.

Here, we present a concrete and transparent mechanism that enforces exceptional deficiency via lattice sum rules 
in non-Hermitian 
square-root topological insulators. 
The square-root construction principle  \cite{Arkinstall2017} provides concrete guidance for the design and interpretation of lattice systems with topological properties, including higher-order topological systems such as quadrupole insulators (QIs) \cite{multipole1,multipole2},  which we will employ to illustrate our results. Applied to exceptional deficiency, 
this approach offers a natural path to identify two subsystems with matching spectra, provided by mutually coupled sublattices, as well as an explicit condition for the required coupling configuration. As square-root topological insulators offer a wide range of phenomena, they also present an ideal platform for identifying signatures of exceptional deficiency. We demonstrate this here for static broadband amplification and adiabatic state amplification, and show that this leads to state-selective and non-Abelian amplification mechanisms. 
Before we discuss these phenomena, we describe the general construction principle of exceptionally deficient topological square-root topological insulators and their implementation in the QI setting.

\begin{figure}[t]
\centering\includegraphics[width=\columnwidth]{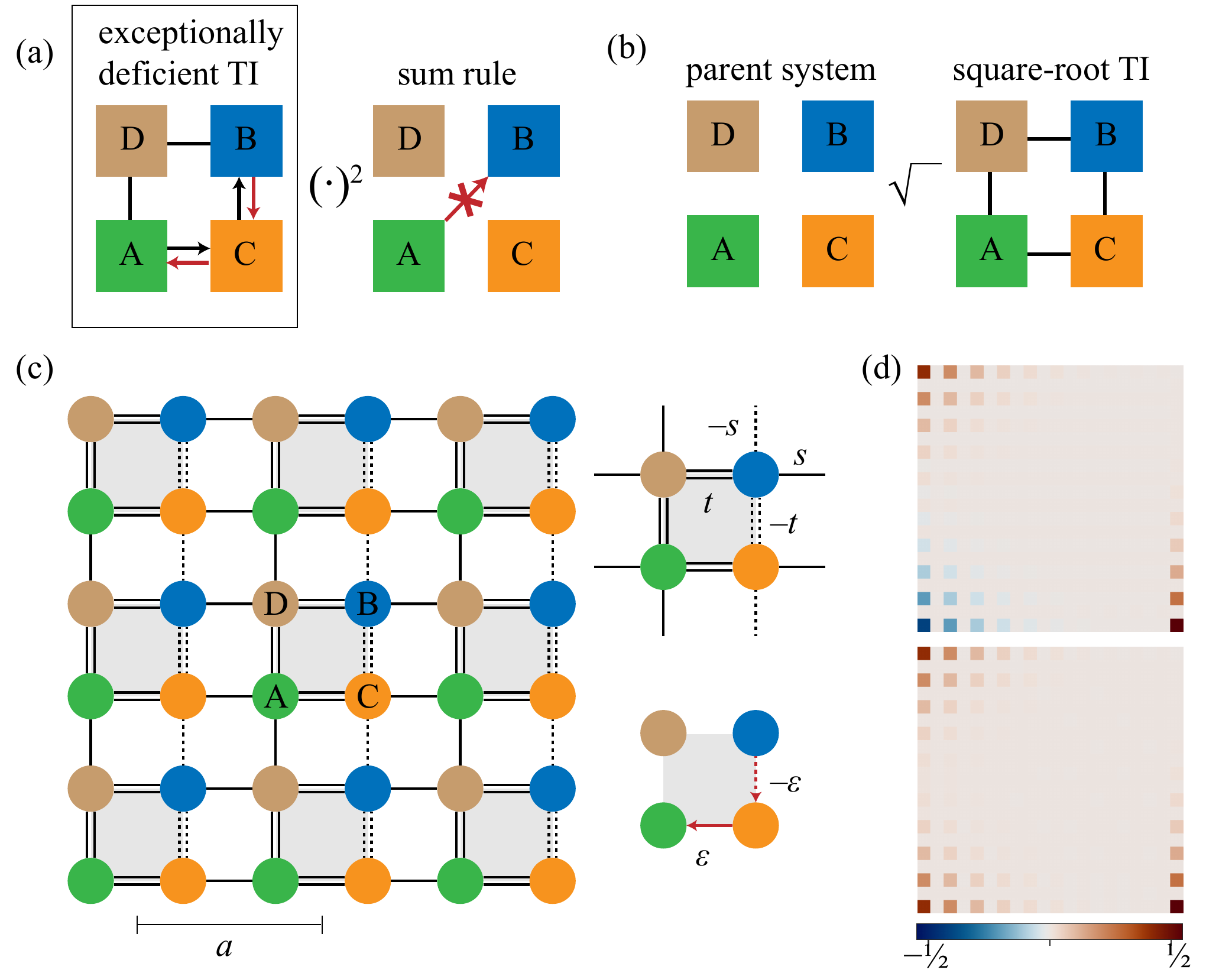}
    \caption{
    (a) Construction principle of an exceptionally deficient topological insulator (TI) based on a sublattice sum rule inherited from a squared parent system. The sum rule constrains the nonreciprocal non-Hermitian couplings with part C (arrows) so that the squared parent system does not couple subpart A to subpart B.  
    (b) Interpretation as a square-root TI, resulting in a nontrivial system with chiral symmetry.      
    (c) Implementation based on a $\pi$-flux  quadrupole insulator with reciprocal couplings $s\equiv 1$ and $t$ as well as additional nonreciprocal couplings of strength $\varepsilon$. (d) The two corner states for the system in (c) with  $10\times 10$ unit cells and in $\varepsilon=-t=1/2$. Each state belongs to two degenerate eigenvalues forming an exceptional point, and this also applies to all other states of this exceptionally deficient system.}
    \label{fig1}
\end{figure}

\emph{Construction principle.}
Figure~\ref{fig1} depicts the general
construction principle of exceptionally deficient topological square-root insulators, along with a specific implementation based on a quadrupole insulator. 
The square-root construction connects uncoupled parent systems with Hamiltonian $H^2$ with a non-trivially coupled system with Hamiltonian $H$. In our example, 
the parent systems provide four sublattices, labelled A, B, C, and D, which are taken to be equivalent in the Hermitian limit. 
We consider nontrivial square roots that mutually couple sublattices (A,B) to sublattices (C,D), resulting in a generic block Hamiltonian
\begin{equation}
H=    \left(
\begin{array}{cccc}
 0 & 0 & H_{AC} & H_{AD} \\
 0 & 0 & H_{BC} & H_{BD} \\
 H_{CA} & H_{CB} & 0 & 0\\
 H_{DA} & H_{DB}& 0 & 0 \\
\end{array}
\right)
\end{equation}
possessing a
chiral sublattice symmetry $\mathcal{X} H \mathcal{X}=-H$ with $\mathcal{X}=\diag(\openone,\openone,-\openone,-\openone)$. Taking the couplings to be real, we also obtain a conventional time-reversal symmetry 
corresponding to complex conjugation $\mathcal{K}$.
Exceptional deficiency will be enforced by demanding the lattice sum rule (lattice assignment without loss of generality)
\begin{equation}
H_{BC}H_{CA}+H_{BD}H_{DA}=0,
\label{eq:sumrule}
\end{equation}
and supplementing this with a generalized transposition symmetry \cite{schomerus2013from,Okuma2020}
$RH^TR=H$, where the involution $R$, inherited from the sublattice permutation symmetry of the parent system, commutes with the chiral operator, $\mathcal{X}R\mathcal{X}=R$. 

Given the sublattice sum rule \eqref{eq:sumrule}, we can construct right eigenstates $\mathbf{u}_l=(\mathbf{a}_l,\mathbf{0},\mathbf{c}_l,\mathbf{d}_l)^T $
 (a column vector) by solving the reduced eigenvalue problem
\begin{equation}
(H_{AC}H_{CA}+H_{AD}H_{DA})\mathbf{a}_l=E_l^2\mathbf{a}_l
\label{eq:red1}
\end{equation}
associated with the A sublattice of the parent system, 
choosing $E_l$ as the positive or negative square root of $E_l^2$,
and completing the eigenstate with
$\mathbf{c}_l=E_l^{-1}H_{CA}\mathbf{a}_l$,
$\mathbf{d}_l=E_l^{-1}H_{DA}\mathbf{a}_l$.
Analogously, we obtain left eigenstates
 $\mathbf{v}_m=(\mathbf{0},\mathbf{b}'_m,\mathbf{c}'_m,\mathbf{d}'_m)$ (a row vector)
by solving the reduced eigenvalue problem
\begin{equation}
\mathbf{b}'_m(H_{BC}H_{CB}+H_{BD}H_{DB})=E_m^2\mathbf{b}'_m
\label{eq:red2}
\end{equation}
associated with the B sublattice of the parent system, 
choosing $E_m$ as the positive or negative square root of $E_m^2$,
and completing the eigenstate with
$\mathbf{c}'_m=E_m^{-1}\mathbf{b}'_lH_{CA}$,
$\mathbf{d}'_m=E_m^{-1}\mathbf{b}'_lH_{DA}$.
The sublattice sum rule \eqref{eq:sumrule} guarantees that 
\begin{equation}
\mathbf{v}_m\cdot\mathbf{u}_l=0,
\label{eq:selforthog}
\end{equation}
and crucially this holds even when $E_l=E_m$. The generalized transposition symmetry transforms the reduced eigenvalue problems \eqref{eq:red1} and \eqref{eq:red2} into each other, so that the constructed right and left eigenstates can be paired up throughout the whole spectrum. In a Hermitian setting, where right and left eigenstates can furthermore be translated into each other, we have constructed a basis, and realize that all eigenvalues are even-fold degenerate, without resorting, for instance, to an explicit Kramers degeneracy \cite{Beenakker1997,Esaki2011}. 
In generic non-Hermitian systems, however, the condition \eqref{eq:selforthog} amounts to self-orthogonality, which marks out EPs \cite{kato,Dembowski2001,Heiss_2004,Heiss_2012,Berry2004}. 
Thereby, we have constructed an exceptionally deficient system.

\emph{QI based realization}.
To illuminate how this construction results in exceptional deficiency, we turn to the concrete implementation given in Fig.~\ref{fig1}(b). This implementation is based on a standard Hermitian QI model \cite{multipole1}, formed by
a unit cell with four sites, intracell couplings $t$, and intercell couplings $s\equiv 1$.
The non-Hermitian modification is obtained from additional nonreciprocal intracell couplings of strength $\varepsilon$, while
the topologically nontrivial features arise from
$\pi$ fluxes attached to each plaquette. The Hermitian system possesses a four-fold
rotational symmetry induced by
\begin{equation}
    C_4=\left(
\begin{array}{cc}
 0 & \openone  \\
 -i\sigma_2 & 0 
\end{array}
\right),
\label{eq:c4}
\end{equation}
with block Pauli matrices $\sigma_i$,
where notably $C_4^4=-\openone$.
The symmetries can further be combined into a generalized time-reversal symmetry operation $\mathcal{T}=\mathcal{K}C_4^2$, $\mathcal{T}^2=-1$, rendering all energy levels in the Hermitian model two-fold Kramers degenerate. Importantly, the $C_4$ symmetry is broken in the non-reciprocal couplings of strength $\varepsilon$, and the system does not display non-Hermitian Kramers degeneracy \cite{Esaki2011}. Instead, the stated sum rule and generalized transposition symmetry enforce exceptional deficiency.

\begin{figure}[t]
\centering\includegraphics[width=\columnwidth]{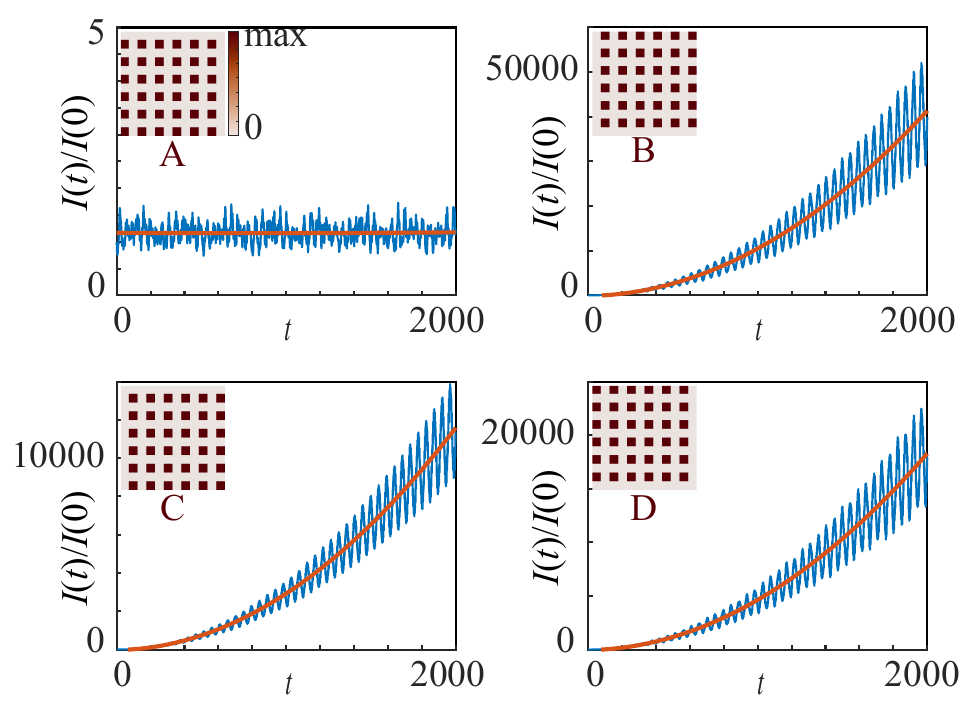}
    \caption{Static state amplification  in the exceptionally deficient QI  with $6\times 6$ unit cells and  fixed $\varepsilon=1/4$, $t=-1/2$. 
    The insets show the intensity distribution of the chosen initial states, which are uniformly localized on the A, B, C, or D sublattices. The main panels display the time dependence of the total intensity.
    Significant state amplification $\propto t^2$  (red curves from polynomial fits) occurs whenever the initial state has components outside of the span of the eigenstates (the A sublattice lies within this span). This provides a broadband signature of exceptional deficiency. 
    }
    \label{figstatic}
\end{figure}

\begin{figure*}[t]
\centering\includegraphics[width=0.9\linewidth]{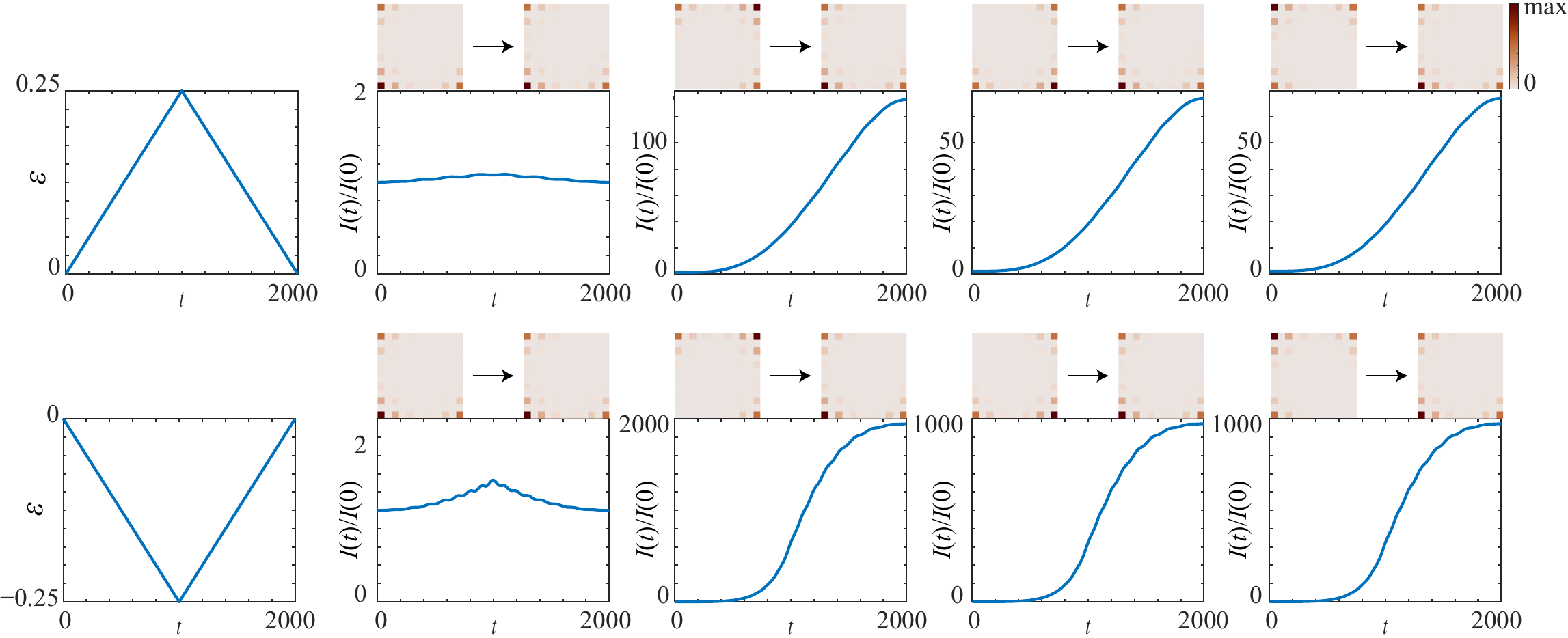}
    \caption{Adiabatic state amplification in the exceptionally deficient QI  with $6\times 6$ unit cells and $t=-1/2$, as $\varepsilon$ is ramped up and down  (top panels) or down and up (bottom panels). The insets show the chosen initial state (predominantly localized in one of the four corners) and the corresponding final state, which is always predominantly localized in the lower left corner. 
    Significant state amplification $\propto (d\varepsilon/dt)^{-2}$ occurs whenever the initial state differs from this dynamically selected final state. This is a signature of the non-Abelian geometry induced by the exceptional deficiency. 
    }
    \label{fig2}
\end{figure*}

By construction of the model, these features apply to finite systems with open boundary conditions. In the range $|t|<1$ the Hermitian QI then exhibits four corner states that appear in two-fold degenerate pairs at opposite energies close to zero. In the non-Hermitian model, these states instead form two EPs. We illustrate these corner states in Fig.~\ref{fig1} (c), where $\varepsilon=-t=1/2$. Because of the exceptional deficiency, not only these corner states, but all size-quantized states in this finite system, are EPs.
Next, we identify two dynamical signatures of the exceptional deficiency in such finite systems, obtained from the state evolution $i\frac{d}{dt}\boldsymbol{\psi}(t)=H(t)\boldsymbol {\psi}(t)$ with static or adiabatically time-dependent Hamiltonian $H(t)$.

\emph{Static broadband state amplification}.
For time-independent Hamiltonians $H$, 
the key dynamical signature of exceptional points is the occurrence of a solution $\exp(-iE_l t)\mathbf{w}_l-it\exp(-iE_l t)\mathbf{u}_l$ along with the standard solution $\exp(-iE_l t)\mathbf{u}_l$, where 
 $E_l$ is the  degenerate eigenvalue,  $\mathbf{u}_l$ the degenerate eigenvector, and
$\mathbf{w}_l$ the generalized eigenvector obeying $H\mathbf{w}_l=E_l\mathbf{w}_l+\mathbf{u}_l$ \cite{kato}.
The total intensity $I(t)=|\boldsymbol{\psi}(t)|^2$ then acquires a $t^2$ dependence, which is absent only when the initial state does not overlap with the generalized eigenvector.
In an exceptionally deficient system,
this static state amplification should occur for any initial condition that lies outside of the span of the eigenvectors $\mathbf{u}_l$, elevating it to a broadband effect.  
To obtain a clear signature of this broadband effect, we utilize the fact that in the QI realization, the A sublattice lies completely within the span of the eigenvectors, while the B sublattice lies completely outside this span.  
As shown in Fig.~\ref{figstatic}, the dynamical signatures are then directly observable in the time-dependent intensity from initial states that are uniformly distributed on a given sublattice. As expected, state amplification is absent for the initial state on the A sublattice, but occurs for
initial states localized on any of the three other sublattices, and it is most
pronounced for the initial state on the B sublattice.

\emph{Non-Abelian adiabatic state amplification}.
Next, we utilize the 
corner states in the QI model to illuminate the dynamical consequences of the exceptional deficiency in a second setting. For this, we exploit the fact that the deficiency is maintained across the whole parameter range, which provides us with the opportunity to explore how states dynamically evolve while being stabilized at an EP.
This feature leads us to consider the phenomenon of adiabatic state amplification \cite{Silberstein2020,Singhal2023,Ozawa2025}, concerning the intensity change $I(t)=|\psi(t)|^2$ of a state as parameters $\boldsymbol\lambda$ are slowly changed along a path $C$. 
This effect is manifestly non-Hermitian, as Hermitian evolution preserves the norm of a state. 
Initialized in a long-living, spectrally isolated, right eigenstate $|R(\boldsymbol\lambda(0))\rangle$, with the long lifetime required so that the adiabatic theorem applies \cite{Nenciu:1992,Graefe:2013}, the state follows the instantaneous right eigenstate $|R(\boldsymbol\lambda)\rangle$, while the
intensity acquires a geometric amplification factor
\begin{equation}
    A_g (C) 
    = 
    \exp
    \left[ - 2\int_{\mathcal{C}} \mathrm{Im}\left[ \boldsymbol{\mathcal{A}}^{LR} (\boldsymbol\lambda) - \boldsymbol{\mathcal{A}}^{RR} (\boldsymbol\lambda)\right]\cdot d\boldsymbol\lambda \right] \label{eq:defgaf}
\end{equation}
connected with the imaginary part 
of the Berry connections
\begin{align}
    \boldsymbol{\mathcal{A}}^{LR} (\boldsymbol\lambda)
    =
    i\frac{\langle L (\boldsymbol\lambda) | \nabla_{\boldsymbol\lambda} R(\boldsymbol\lambda)\rangle}{\langle L (\boldsymbol\lambda)|R(\boldsymbol\lambda)\rangle},
    \label{eq:alr}   \\
    \boldsymbol{\mathcal{A}}^{RR} (\boldsymbol\lambda)
    =
    i\frac{\langle R (\boldsymbol\lambda) | \nabla_{\boldsymbol\lambda} R(\boldsymbol\lambda)\rangle}{\langle R (\boldsymbol\lambda)|R(\boldsymbol\lambda)\rangle} \label{eq:arr},
\end{align}
where $\langle L (\boldsymbol\lambda)|$ is the associated instantaneous left eigenstate. 
As the intensity is a measureable physical quantity, the geometric amplification factor is gauge invariant even when the path  $C$  is open. On the other hand,
expression \eqref{eq:defgaf} implies $A_g=1$ for any closed path $C$ that retraces itself back to the starting point, a feature tied to the Abelian nature of the associated complex Berry phase.

The Berry phase becomes non-Abelian when the state is degenerate \cite{Wilczek1984}, and in the non-Hermitian setting, this non-Abelian phase again acquires imaginary parts \cite{snizhko2019non2}. As shown in Fig.~\ref{fig2}, concrete non-Abelian effects manifest for the exceptionally deficient non-Hermitian QI, making them directly observable in the state amplification. In the figure, we amplify a state by slowly varying $\varepsilon$ over time from the Hermitian starting point $\varepsilon=0$ to  $\varepsilon=\pm 1/4$, and then back to  $\varepsilon=0$. For the initial state we consider, in turn, one of the four corner states, predominantly localized on the A, B, C, or D sublattice (see insets). The first non-Abelian manifestation is the observation that at the end of the evolution, the final state has been transformed into the state predominately localized on the A sublattice. The second non-Abelian manifestation is the observation that, with the exception of the situation where the initial state is equal to this universal state, a significant intensity amplification has occurred even though the path $C$ retraced itself.  
The state amplification is systematic, 
scaling as $\sim (d\varepsilon/dt)^{-2}$ \cite{Holler:2020} in analogy with the $t^2$ scaling of static amplification described above, and
is equal for the two evolutions starting predominantly on the C and D sublattices, while the amplification of the state starting on the B sublattice is twice as large. 

\emph{Signatures of broken bulk-boundary correspondence.} 
The dynamical effects described so far occur for finite systems, benefiting from the exactness of the described construction principle.
However, the spectral features of non-Hermitian systems with non-reciprocal couplings are known to be highly sensitive to the boundary conditions, invalidating the bulk-boundary principle that connects finite and infinite periodic systems in Hermitian topology \cite{RevModPhys.82.3045,RevModPhys.83.1057}. This characteristic is exemplified by the non-Hermitian skin effect \cite{Yao2018,Kunst2018,Yokomizo2019,Lee2019,Zhang2020}, where the bulk and boundary spectra for periodic and open boundary conditions drastically differ. In the described non-Hermitian QI model,
we find that while states of the infinitely periodic system remain strictly degenerate, the degeneracies are no longer necessarily exceptional points, so that the system is only partially exceptionally deficient. This behavior follows directly from analyzing the Bloch Hamiltonian 
\begin{align}
H(k_x,k_y)=
\left(
\begin{array}{cc@{\!\!\!\!\!}cc}
 0 & 0 & r+e^{-ik_x}& t+e^{-ik_y} \\
 0 & 0 & - t- e^{ik_y}   & t+e^{ik_x}  \\
 t+e^{ik_x} & - r- e^{-ik_y} & 0 & 0\\
 t+e^{ik_y} & t+e^{-ik_x} & 0 & 0 \\
\end{array}
\right)
,
\end{align}
where $r=t+\varepsilon$ and we set the lattice constant $a\equiv 1$.
The sum rule \eqref{eq:sumrule} is obtained from the identity
\begin{equation}
    (- t- e^{ik_y})(t+ e^{ik_x})+(t+ e^{ik_x})(t+  e^{ik_y})=0,
\end{equation}
highlighting the role of the $\pi$ fluxes generated by the nontrivial square root. 
The generalized transposition symmetry takes the form
\begin{equation}
    \left(
\begin{array}{cc}
 \sigma_1 & 0  \\
0 & -\sigma_3\\
\end{array}
\right)H(k_x,k_y)   \left(
\begin{array}{cc}
 \sigma_1 & 0  \\
0 & -\sigma_3\\
\end{array}
\right)=H^T(k_y,k_x), 
\end{equation}
amounting to a reflection about the line $x+y=0$. 
Hence, the four bands 
\begin{align}
&E^{(1)}_\pm(k_x,k_y)=\pm\sqrt{|t+e^{ik_x}|^2+|t+e^{ik_y}|^2+
\varepsilon(t+e^{ik_x})
},
\nonumber
\\
&E^{(2)}_\pm(k_x,k_y)=\pm\sqrt{|t+e^{ik_x}|^2+|t+e^{ik_y}|^2+
\varepsilon(t+e^{ik_y})
},
\label{eq:bands}
\end{align}
whose winding in the complex plane signifies the non-Hermitian skin effect \cite{Zhang2020},
respect the spectral symmetry $E^{(1)}_\pm(k_x,k_y)=E^{(2)}_\pm(k_y,k_x)$.
For $k_x\neq k_y$, the band degeneracy occurs between different locations in the Brillouin zone, involving two mutually orthogonal extended Bloch states, as in a Kramers-degenerate system. 
Therefore, the continuous spectrum of this system exhibits exceptional deficiency only for $k_x=k_y\equiv k$, where each of the twofold degenerate bands
$E^{(1)}_+(k,k)=E^{(2)}_+(k,k)$
and
$E^{(1)}_-(k,k)=E^{(2)}_-(k,k)$
is associated with a unique extended Bloch eigenstate $\mathbf{u}_\pm(k,k)e^{ik(x+y)}$.
The proposed construction principle of exceptionally deficient models thus also opens up a new avenue for exploring the breakdown of the 
bulk-boundary correspondence in non-Hermitian systems.

\emph{Conclusions and outlook.} 
In summary, we provided a transparent construction principle to design non-Hermitian systems that exhibit exceptional deficiency. The square root construction allowed us to equip the system with topological features, which we exemplified by a quadrupole insulator model with corner states. Exceptional deficiency gives rise to striking dynamical signatures,
including in broadband and adiabatic state amplification, and unveils non-Abelian geometric characteristics that can be utilized for mode conversion.
Our framework is readily implementable in existing platforms such as topolectric circuits \cite{helbig2020generalized,Zou2021}, active acoustic and mechanical metamaterials \cite{brandenbourger2019non,ghatak2020observation,Zhang2021,veenstra2024non}, and quantum-optical systems \cite{Wanjura2020}, including dissipative cold atom setups \cite{liang2022dynamic}.
Thereby, our work establishes a versatile route for designing exceptionally deficient systems 
that can be flexibly equipped with desired features, enabling new phenomena across a broad range of non-Hermitian settings.

This research was funded by EPSRC via Grant No.
EP/W524438/1.

The data that support the findings of this work are openly
available \footnote{Research datasets at \url{https://doi.org/10.5281/zenodo.16540598}}.


%

\end{document}